\documentclass[twocolumn,twoside,slac_two]{revtex4}
\usepackage{graphicx}
\usepackage{fancyhdr}
\begin{document}

\title{Status and recent results from H.E.S.S.}

%

\author{B. Giebels, for the H.E.S.S. Collaboration}
\affiliation{Laboratoire Leprince-Ringuet, Ecole polytechnique IN2P3/CNRS, Route de Saclay, F-91128 Palaiseau}

\begin{abstract}
  The H.E.S.S. instrument consists of four $13\,{\rm m}$ (H.E.S.S. I) and one $28\,{\rm m}$ diameter
  (H.E.S.S. II) Atmospheric Cherenkov Telescopes (ACTs) located in the Khomas Highland in Namibia, 1800 m
  above sea level. The H.E.S.S. I array began operations in 2003, and has achieved recent scientific results
  allowed by the synergy with the {\it Fermi}-LAT, some of which are outlined here. The H.E.S.S. II telescope
  started operations in July 2012, and is expected to provide its first scientific results by the end of the
  year. Since the inauguration of the first telescope in September 2002, H.E.S.S. has taken 9415 hours of
  data, with 4234 hours in the band of the Galaxy and 5181 hours in extragalactic space, discovered over 80 new very
  high energy (VHE;$E > 100\,{\rm GeV}$) $\gamma$-ray sources (according to TeVCat listings), among them more
  than 60 galactic objects and 19 extragalactic sources.

\end{abstract}

\maketitle

\thispagestyle{fancy}


\section{The H.E.S.S. Array}

\vspace{-2ex}

{\bf With its 28-meter-sized segmented mirror, H.E.S.S. II is the largest ACT ever built}. The peak
positioning speed of 100 degrees/minute allows to point the instrument extremely fast towards transient
phenomena such as $\gamma$-ray bursts, and its large mirror allows the detection of sources down to a few tens
of GeV. The combination of all 5 ACTs in hybrid mode allows an improved sensitivity thanks to better defined
shower images.  H.E.S.S. II saw its first light at 0:43 a.m. on 26 July 2012. The {\bf H.E.S.S. I telescopes}
are undergoing a long process of different upgrades. Recently, all 380 mirrors of each $13\,{\rm m}$
tesselated telescope mirrors were recoated over the span of 2 years, making for a spectacular recovery of
optical efficiency which dropped progressively over the past $\sim$8 years of operations. More improvement is
expected when the ageing Winston cones, phototubes and electronics will be replaced as well.

Thanks to the improved shower image quality and stereoscopy, the reconstruction techniques have also improved
beyond the standard Hillas parametrization. Reconstruction methods using all pixels in cascade fits, 3-dimensional
characterizations,  or the use of multivariate methods such as boosted decision trees, have improved by a factor
$\sim$2 the flux sensitivity of the H.E.S.S. analysis (see references in \citealt{gamma2012}).

The expected sensitivity of the H.E.S.S. array including the H.E.S.S. II telescope should result in an
improvement in the 100 GeV range as well as the decrease in energy threshold. This should allow a more
performant search for pulsed emission in some Galactic sources, improve the chances to catch the elusive VHE
$\gamma$-ray glow GRBs, as well as detect new classes and more distant extragalactic objects.

\section{Galactic emitters}

\begin{figure}
\includegraphics[width=65mm]{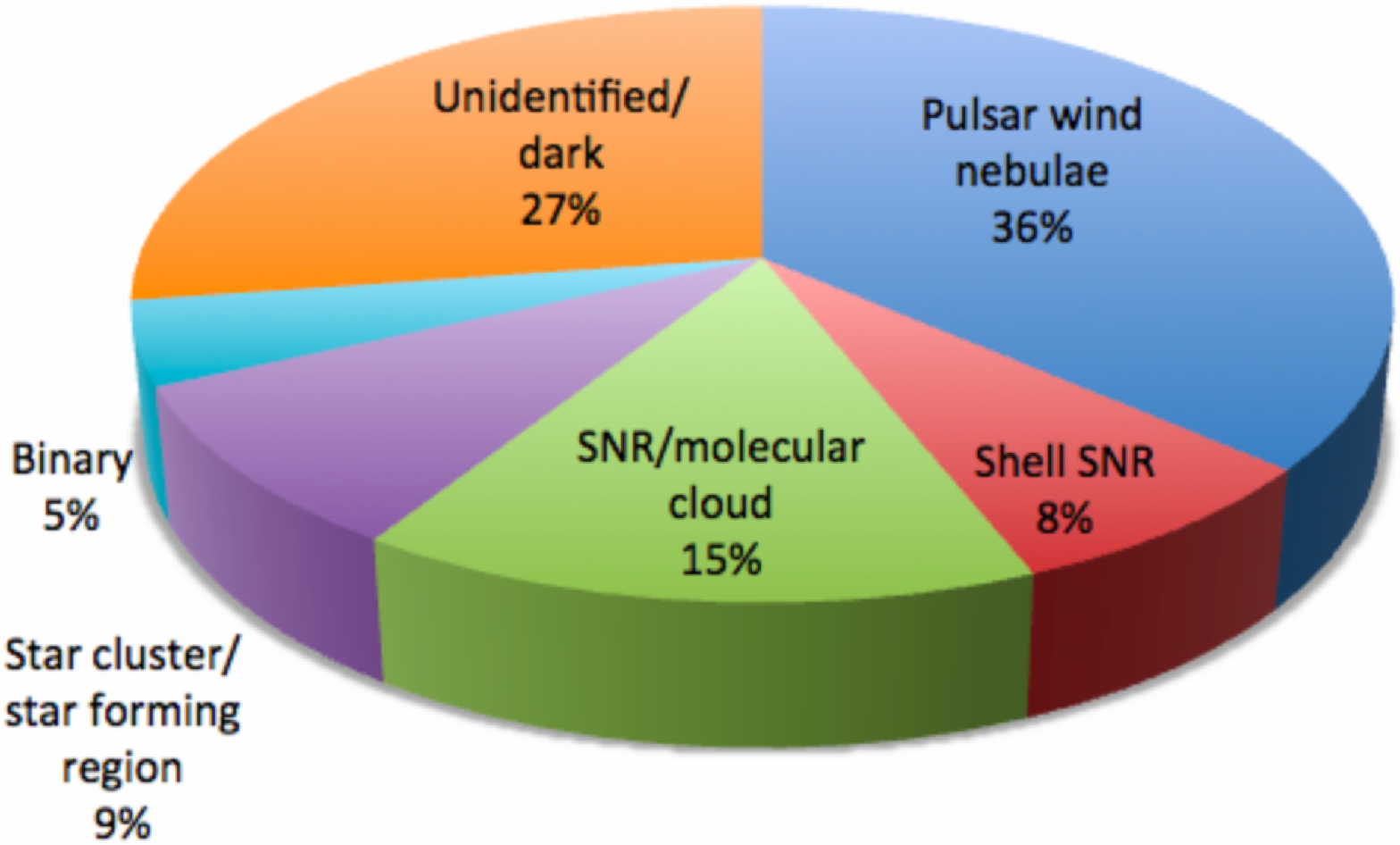}
\caption{Fraction of the different type of VHE $\gamma$-ray emitters revealed from the  H.E.S.S. Galactic Plane Survey.}
\label{f1}
\end{figure}

Most rewarding in terms of source discoveries proved to be the {\bf H.E.S.S. Galactic Plane Survey}, which revealed
a large variety of sources of VHE $\gamma$-rays lining the Milky Way. Pulsar wind nebulae (PWNe) - giant
bubbles filled with electrons and positrons created by spinning neutron stars - emerge as the most abundant
source type. The breakdown of nature of the Galactic sources, as classified in {\tt http://tevcat.in2p3.fr} , is shown in Figure~\ref{f1} illustrating the diversity of celestial $\gamma$-ray
emitters. Most likely, a significant fraction of the unidentified sources are PWNe, where the
pulsar is not (yet) detected. Work is ongoing to provide flux maps based on the H.E.S.S. Galactic Plane Survey as
well as a unified source catalog based on a semiautomatic pipeline, as a basis for ensemble studies of source
classes \citep{icrc2011}.

{\bf Resolved supernova remnant shells} - the 'classical' cosmic particle accelerators - as well as {\bf supernovae
interacting with molecular clouds, binary systems, and star clusters}, constitute the next most abundant type
of $\gamma$-ray sources in the Galactic Plane. The point in case here is the W49 region, a prime candidate for ACT
observations since it hosts a star forming region (W49A) and a mixed morphology supernova remnant interacting
with molecular clouds (W49B).

\begin{figure}
\includegraphics[width=65mm]{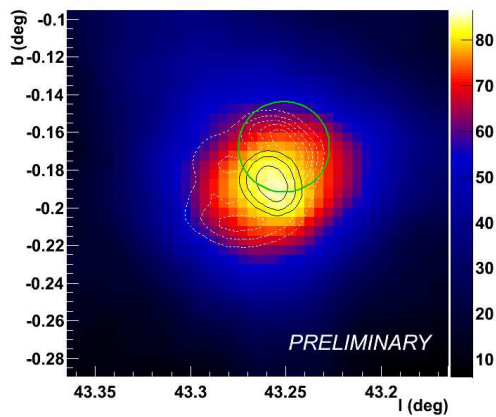}
\caption{Skymap of W49B with black error contours at 68\%, 95\% and 99\%
of the fitted position assuming point-like emission. The green circle is the {\it Fermi}-LAT position at 95\% C.L. The
white contours show the radio emission as seen by NVSS. From \cite{brun}}
\label{f2}
\end{figure}

Figure~\ref{f2} shows the H.E.S.S. $\gamma$-ray excess towards W49B, well coincident with the brightest radio
emission of the W49B remnant (white contours), and with the GeV emission detected by the {\it Fermi}-LAT (green
circle). The combined GeV-TeV spectrum shows a smooth connection between both regimes. Given the
very high GeV luminosity and the fact that the SNR is interacting with dense material, a hadronic scenario is
favored \citep{brun}.

A few of the {\bf binary systems} observed by H.E.S.S. are particularly noteworthy here.

{\it (i)} Observations of the field of view around the position of the new $\gamma$-ray binary
discovered by {\it Fermi}-LAT, 1FGLJ1018-5856, shows the presence of HESSJ1018-586, a combination of a point-like source
which could be the VHE counterpart of the binary system, and an extended emission region probably related
to the nearby pulsar PSRJ1016-5857. The spectral analysis shows a photon index $\Gamma=2.7\pm0.5$ for the
point-like source, with a normalization at 1 TeV of $(4.2\pm1.1)\times 10^{-12}\,{\rm TeV}^{-1}\,{\rm
  cm}^{-2}\,{\rm s}^{-1}$. Inspection of the source
lightcurve does not reveal any clear indication of periodicity, although further H.E.S.S. Observations
homogeneously sampling the phase-space range are required to properly asses the source variability at TeV
energies. 

\begin{figure}
\includegraphics[width=85mm]{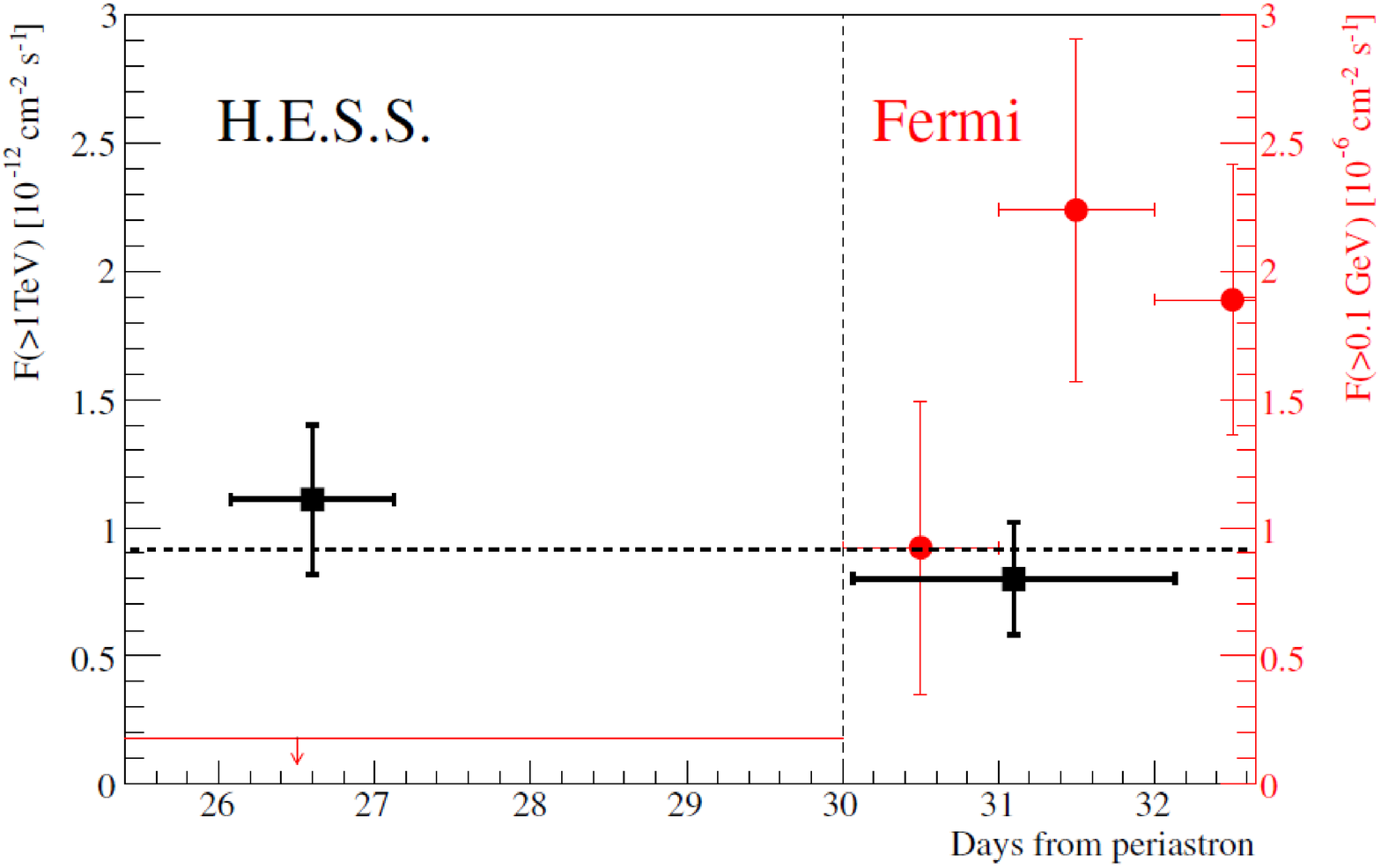}
\caption{The integrated photon fluxes for H.E.S.S. and {\it Fermi}-LAT for the pre-flare and flare periods . The
  dashed horizontal line shows the best fit with a constant (see also \citealt{lurii}). }
\label{f3}
\end{figure}

{\it (ii)} The binary system PSR B1259-63 was observed in the 2010/2011 periastron passage, in which
  contemporaneous {\it Fermi}-LAT data were also taken, showing an unexpected flare occuring about 30 days
  after periastron. Whilst the GeV flare increases by a factor {$\geq 9.2$} between the pre-flare and flare flux
  levels, at VHE the flux does not increase by a factor greater than 3.3 at the 99.7\% confidence level
  (Fig.~\ref{f3}). This implies that the GeV flare and the TeV emission have a different physical origin (see
  also \citealt{lurii,dubus}).

{\it (iii)} The field of view around $\eta$ Carinae and the Carina nebula was observed during 2004-2010. The
massive binary system $\eta$ Car is of particular interest due to its spatial coincidence with the {\it
  Fermi}-LAT source 2FGLJ104-.05941. The H.E.S.S. observations account for about 33.1 h live-time and
corresponds to six periods covering different orbital phases. No detection is found at energies $>470\,{\rm
  GeV}$ yielding upper limits to the emission from the central system and the extended Nebula at fluxes $0.72\times
10^{-12}\,{\rm cm}^{-2}\,{\rm s}^{-1}$ and $4.4\times 10^{-12}\,{\rm cm}^{-2}\,{\rm s}^{-1}$,
respectively. If the {\it Fermi}-LAT spectrum of $\eta$ Car is extended to the VHE regime, it would have been
detectable in the current H.E.S.S. data. Therefore the non-detection of a significant VHE $\gamma$-ray signal from
$\eta$ Car in the complete H.E.S.S. data set implies the presence of a spectral cut-off in the range 0.1-1
TeV.

HESS J1646-458 is a new VHE $\gamma$-ray source found towards the unique massive {\bf stellar cluster}
Westerlund 1 with observations performed in 2004, 2007-2008 for a total 33.8 h live-time \citep{wester}. The spectrum in the
range 0.45-75 TeV of the entire region is well fit by a power law with $\Gamma=2.19\pm0.08$ and a
normalisation at 1 TeV of $(9.0\pm1.4) \times 10^{-12}\,{\rm TeV}^{-1}\,{\rm cm}^{-2}\,{\rm s}^{-1}$. The
integrated flux above 0.2 TeV amounts to $(3.49\pm0.52)\times 10^{-11}\,{\rm cm}^{-2}\,{\rm s}^{-1}$ . The VHE
$\gamma$-ray luminosity between 0.1-100 TeV is $1.9\times1035 (d/4.3 {\rm kpc})^2\,{\rm erg}\,{\rm
  s}^{-1}$. The large size of HESS J1646-458, over $2^\circ$ in diameter, makes it one of the largest VHE
sources so far detected by H.E.S.S. and presents a challenge in identifying clear counterpart(s) to explain
the VHE $\gamma$-ray emission. Among the potential counterparts are three unidentified {\it Fermi}-LAT sources
2FGLJ1650.6-4603c, 2FGLJ1651.8-4439c, and 2FGLJ1653.9-4627c, as well as the $\gamma$-ray pulsar
2FGLJ1648.4-4612. The latter has been associated with the high spin-down power pulsar PSR J1648-4611, although
the unprecedented high conversion efficiency needed and the size of the VHE emission region disfavor HESS
J1646-458 as a single, very extended PWN.

\vspace{-7ex}

\section{Extragalactic sources}
\vspace{-3ex}
The small field of view (a few degrees) and the limited number of observation hours ($\sim$1000h/yr) from ACTs
require elaborate strategies to optimize the detection of new {\bf extragalactic VHE emitters}. With the {\it
  Fermi}-LAT effective area improvement over EGRET's at energies $E>10\,{\rm GeV}$, the successive $\gamma$-ray
catalogs have been an unvaluable tool to direct observations from all currently operating ACTs. Even in the
case of BL Lac objects, of which only 50\% have a firmly established redshift, a {\it Fermi}-LAT spectrum with
a photon index in the range $1.5<\Gamma<2$ and a few photons at energies $\sim 100\,{\rm GeV}$ are strong
indicators of potentially successful VHE measurements. This is the case for the recently detected BL Lac
objects PKS 0447-439 and, more recently, of the object KUV 00311-1938 which, at $z\sim0.6$, is currently the most
distant source of VHE $\gamma$-rays.

\begin{figure}
\includegraphics[width=85mm]{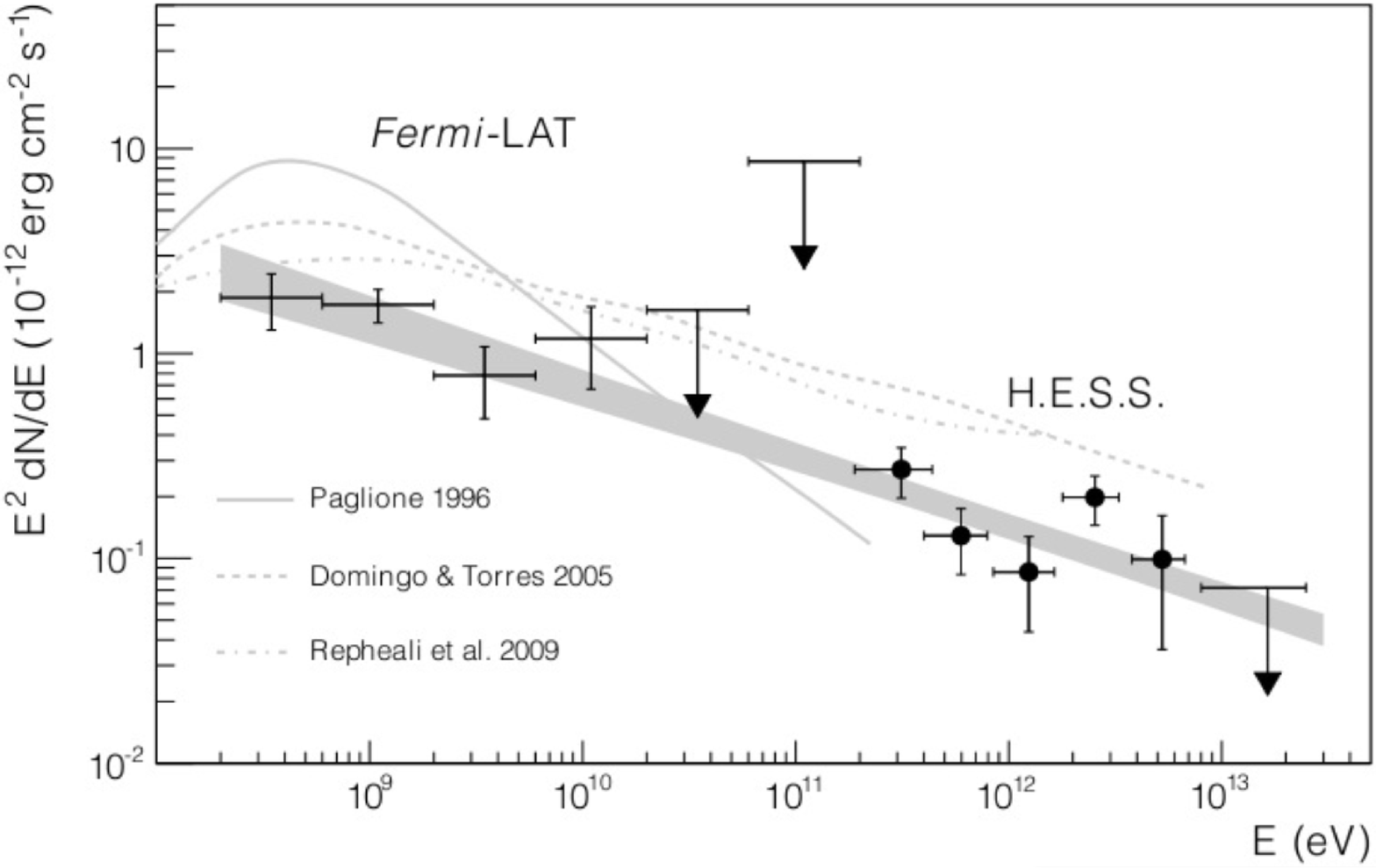}
\caption{Differential H.E.S.S. and {\it Fermi}-LAT energy spectrum, with theoretical predictions, described
  further in \cite{ngc}.}
\label{f5}
\end{figure}

The $\gamma$-ray observations performed with H.E.S.S. and {\it Fermi}-LAT allow also to investigate the
non-thermal processes in the {\bf starburst galaxy} NGC 253. The $\gamma$-ray source is compatible with the
optical centre of NGC 253. The VHE $\gamma$-ray data can be described by a power law in energy with
$\Gamma=2.14$ and differential flux normalization at 1 TeV of $\sim 9.6\times10^{-14}\,{\rm TeV}^{-1}\,{\rm
  cm}^{-2}\,{\rm s}^{-1}$ . A power-law fit to the {\it Fermi}-LAT spectrum reveals $\Gamma=2.24$ and an
integral flux between 200 MeV and 200 GeV of $\sim 4.9\times 10^{-9}\,{\rm cm}^{-2}\,{\rm s}^{-1}$. No
evidence for a spectral break or turnover is found over the dynamic range of both the LAT instrument and the
H.E.S.S. experiment: a combined fit of a power law to the {\it Fermi}-LAT and VHE $\gamma$-ray data results in
a $\Gamma=2.34$ with a p-value of 30\%. The $\gamma$-ray observations indicate that about 20\% - 30\% of the
cosmic-ray (CRs) energy of the capable of producing hadronic interactions is channeled into pion production. The
smooth alignment between the spectra in the {\it Fermi}-LAT and VHE $\gamma$-ray domain (Fig.~\ref{f5})
suggests that the same energy-loss processes dominate in the {\it Fermi}-LAT and H.E.S.S. energy
range. Advection is most likely responsible for charged particle removal from the starburst nucleus from GeV
to multiple TeV energies. In a hadronic scenario for the $\gamma$-ray production, the single overall power-law
spectrum observed would therefore correspond to the mean energy spectrum produced by the ensemble of CR
sources in the starburst region.

\section{Exotic physics and propagation effects}

From 2004 to 2008, H.E.S.S. acquired 112h of observations of the Galactic Centre (GC) region. A particular
observation strategy was used in order to detect VHE radiation from WIMP annihilations, where a
region close to the Galactic Centre was defined, expected to hold a larger {\bf dark matter} (DM) density than
other areas from which the cosmic-ray background was determined. Regions contaminated by known sources
unrelated to DM were excluded. No significant excess of VHE radiation was found, allowing to establish
upper-limits in the DM mass vs velocity weigthed annihilation cross-section plan for two particular models of
DM distribution profile, NFW and Einasto, both consistent with N-body simulations. The
central part of the Galaxy, suffering large uncertainties on the profile adopted by the DM, was excluded from
the analysis. The constraints established by H.E.S.S. from these observations  correspond to the best limits
obtained so far for DM particle masses between few 100 of GeV to few 10 of TeV, and are complementing the
results obtained by the {\it Fermi}-LAT at lower mass range, like the one obtained in direction of the dwarf
spheroidal galaxy Draco \citep{dmref}.

\begin{figure*}[t]
\centering
\includegraphics[width=135mm]{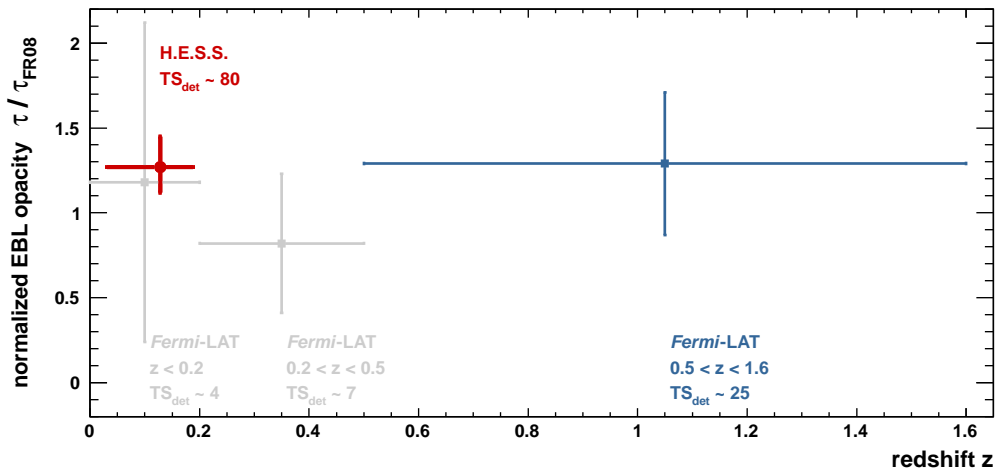}
\caption{The EBL density, as measured by H.E.S.S. and {\it Fermi}-LAT, as a function of distance.} \label{f7}
\end{figure*}


According to different Quantum Gravity theories, {\bf Lorentz symmetry} might be broken at energy scales close
to the Planck scale ($\sim10^{19}\,{\rm GeV}$). In those theories, the light velocity would vary as a function
of the energy of the photons. As a consequence, time delays for photons from cosmological sources showing fast
flux variability (GRBs, flares of AGNs), might be measurable by space or ground based $\gamma$-ray
telescopes. Contrary to space-based telescopes such as {\it Fermi}-LAT, which can observe GeV events from
very distant sources (up to $z\sim8$) but with limited statistics, ACTs such as H.E.S.S. are able to detect
large number of events with energies up to few tens of TeV, but only for nearby sources.  On July 28, 2006, the
H.E.S.S. experiment observed an extreme flare from the AGN PKS~2155-304, located at a redshift of
$z=0.116$. More than 8000 on-source events were recorded during $\sim$85 minutes of observation of this
transient phenomena, for energies $E>120\,{\rm GeV}$. After application of tight constraints on the energy and
direction of the incoming events, and performing an event-by-event likelihood fit on the observed lightcurve
of PKS~2155-304, constraints were derived on the first and second orders of the photon dispersion relation
development. The 95\%CL lower limit on the Quatum Gravity energy scale obtained from the linear,
resp. quadratic, term is $2.1\times10^{18}\,{\rm GeV} (0.5\times10^{11}\,{\rm GeV})$, thus the best limits
obtained so far from AGN studies \citep{liv2011}.


The $\gamma$-rays emitted by sources at cosmological distances can interact with the {\bf Extragalactic Background
  Light} (EBL) through $\gamma \gamma \rightarrow e^+e^-$. Distortions in the observed spectrum can be used to
estimate its density in the $0.1-10\,{\rm \mu m}$ range with VHE $\gamma$-rays and, when sufficient photons statistics is
available, specific wavelength ranges of the EBL can be probed. This yielded the first
direct detection of the presence of EBL using VHE $\gamma$-rays \citep{ebl}. A similar method was used by the
{\it Fermi}-LAT collaboration to probe the EBL density at distances larger than those accessible by Cherenkov
telescopes, which have however larger effective areas and hence a better statistical uncertainty - both
experiments are nevertheless in remarkable agreement when the source distances overlap (Figure~\ref{f7}). {\bf Acknowledgement: please see standard acknowledgement in H.E.S.S. papers,
 not reproduced here due to lack of space}

\bigskip 

\end{document}